# Cerebralization of mathematical quantities and physical features in neural science: a critical evaluation

*Laurent* Goffart[1]

[1]Centre Gilles Gaston Granger, UMR 7304 CNRS Aix Marseille Université, Aix-en-Provence, France

**Abstract.** At the turn of the 20th century, Henri Poincaré explained that geometry is a convention and that the properties of space and time are the properties of our measuring instruments. Intriguingly, numerous contemporary authors argue that space, time and even number are "encoded" within the brain, as a consequence of evolution, adaptation and natural selection. In the neuroscientific study of movement generation, the activity of neurons would "encode" kinematic parameters: when they emit action potentials, neurons would "speak" a language carrying notions of classical mechanics. In this article, we shall explain that the movement of a body segment is the ultimate product of a measurement, a filtered numerical outcome of multiple processes taking place in parallel in the central nervous system and converging on the groups of neurons responsible for muscle contractions. The fact that notions of classical mechanics efficiently describe movements does not imply their implementation in the inner workings of the brain. Their relevance to the question how the brain activity enables one to produce accurate movements is questioned within the framework of the neurophysiology of orienting gaze movements toward a visual target.

## 1 Introduction

Over the past decades, confusion has spread within the domain of cognitive neuroscience between measurement values, the physiological processes generating them and some general physico-mathematical notions. By directly linking their measurement to concepts, some authors imagined the possibility of an internal "coding" of space, time, number (quantity), geometry and kinematics within the brain [1-5]. The suggestion that mathematical and physical notions and practices are cerebralized may look puzzling for those who consider them as human cultural or socio-historical products [6-8], i.e., as collective works of art developed over several generations of intellectuals. For Henri Poincaré, "*the geometric language is after all only a language*" and "*space is only a word that we have believed a thing*" [9]. More crucially, Pellionisz and Llinas explained that the usage of separate space and time coordinates is inapplicable for describing the inner workings of the brain because there is no simultaneity agent comparable to light within the central nervous system [10].

When we examine, for instance, contemporary neurophysiological studies of eye movement generation, we discover kinematic parameters being "encoded" in the activity of some central neurons, in accordance with signals used in modeling studies. By emitting action potentials, some neurons send signals to their post-synaptic targets that would correspond more or less faithfully to the instantaneous angular distance between gaze and target directions, or to the angular velocity of the orienting gaze movement. The measured physical changes in the orientation of the eyes and head replace not only the complex set of contractions and relaxations of extraocular and neck muscles [11-16], but also associated autonomic responses [17]. The multiple processes unfolding in parallel in the brain and converging onto the groups of extraocular and neck motor neurons are considered to be devoted to orienting the eyes or gaze (sum of eyes, head and trunk direction) by reducing mismatches between gaze and target directions (gaze error) or between their velocities (velocity error).

This physico-neurophysiological correspondence is problematic because it has led several authors to introduce the notion of "noise" as an intrinsic part of the brain activity whenever the correlation between the physical measurement values and the firing rate of neurons is low. In the absence of empirical neurophysiological data, also despite evidence to the contrary [18], some authors have posited an "*internal variability in the motor commands*" for generating saccades. They proposed that one part of the brain, the cerebellum would compensate for noisy neuronal signals and account for physically precise movements [19]. Yet, in the context of saccade generation, Sparks proposed that "*the precision or accuracy of a saccade results from the summation of the movement tendencies produced by the population of neurons rather than the discharge of a small number of finely tuned neurons*" and that "*the effects of variability or "noise" in the discharge frequency of a particular neuron... [could be] reduced by averaging over many neurons*" [20]. More fundamentally and as explained elsewhere [21], there is neither reason nor necessity for the firing rate of individual neurons to fit with the dimensionality or the sampling rate of our recording instruments. Neuroanatomical and electrophysiological studies reveal a diversity of neuronal types and distributed connectivity, such that not all neuronal activities lead to the rotation of the eyes. The neuronal events that our instruments and cognitive tools cannot grasp might be more than a mere noisy and meaningless hubbub. In this article, we take advantage of the considerable knowledge gathered during the last six decades, in the neuroanatomy and neurophysiology of the primate oculomotor system [11-17], to question the relevance of general notions belonging to the physical sciences of movement for advancing further our understanding of the inner workings of the brain networks and of its relations to mathematics.

---

* Corresponding author: laurent.goffart@cnrs.fr

## 2 A few words about the movement measurement

Prior to considering whether the firing rate of neurons might encode a kinematic parameter, we shall discuss what the latter is. In most neurophysiological studies in the trained monkey, the orientation of the eye is recorded with a method based upon the electromagnetic induction [22]. The animal is placed in front of a visual display with its head at the center of a cube composed of two orthogonal magnetic fields, one horizontal and one vertical. A coil of thin biocompatible cable previously fixed to one eyeball enables measurement of its angular deviation with respect to the magnetic fields. Most often, the monkey is tested with the head restrained, but a coil fixed to the head enables one to record its changes in orientation as well. Each recording consists of sequences of numerical values that quantify the horizontal and vertical orientations successively taken by the eye. The horizontal orientation corresponds to the amount of angular deviation relative to the body midsagittal plane and along the plane passing through the four extraocular muscles (lateral and medial rectus muscles attached to each eye) responsible for the horizontal movements. The vertical orientation corresponds to the amount of angular deviation relative to the horizontal plane. When the head is upright, the intersection between the horizontal and body midsagittal planes defines the straight-ahead direction. By convention, negative values of horizontal and vertical orientations correspond to leftward and downward deviations and positive values to rightward and upward deviations, respectively. Starting from the straight-ahead direction (for which the zero value is assigned to both horizontal and vertical orientations), a rightward movement is numerically represented by a sequence of positive values ranging from an initial value (corresponding to the movement onset) to a final value (movement end).

A kinematic parameter is a physical magnitude such as the movement amplitude, or a combination of magnitudes as the movement speed (ratio between movement amplitude and duration). It is associated with a continuum of numerical values that are totally ordered. It is a continuum insofar as an infinite number of intermediate ordered values can potentially insert between two successive recorded values. This continuum allows representing the sequence of numerical values by a line segment instead of a sequence of points. Between two successive values, the length of the interval is proportional neither to the amount of effort that was required to change the orientation or the position of the body segment, nor to the amount of energy that was spent. It is proportional to the number of pulses emitted at regular intervals by an external clock. Drawing another line segment (called time) orthogonal to the first one and plotting the numerical values of orientation as a function of the cumulated number of pulses makes it possible to distinguish movements of identical amplitude but with different durations. Further information provided by such a graphical plot is that the time course of orientation values is not rectilinear but slightly curved. The rate of change in eye or head orientation, its angular speed, is not constant. By calculating the ratio between the amplitude and duration of successive changes in orientation and plotting it as a function of time reveals two intervals: a first interval (called acceleration) during which the speed increases to a maximum followed by a second interval (called deceleration) during which it returns to zero. As for the changes in eye or head orientation, the successive speed values are ordered and chained by a continuum of intermediate values. We shall now see that the continuity of measurement values contrasts with the discontinuous character of neuronal activity in the brain.

## 3 A few basic facts about neuronal activity

In brain activity, a radically different structure corresponds to the sequences of numerical values. Let's consider the generation of a horizontal saccadic eye movement. Between the two optic nerves (input side) and the nerves innervating the extra-ocular muscles (output), the connectivity is not reducible to independent lines transmitting signals, for example from ganglion cells aligned along the binocular horizontal plane to motor cells innervating the medial and lateral rectus muscles. Its structure is doubly arborescent, initially diffluent then confluent. The signals are not like a series of values ordered within a continuum either. They consist of sequences of unitary events called action potentials (spikes) separated by a relatively silent interval. The spikes emitted by each neuron propagate from the soma to its pre-synaptic endings along an axon that spatially diverges and contacts multiple post-synaptic neurons. The latter can be located in different groups of neurons and at different distances from the former. Moreover, the speed of propagation is not uniform, but varies between neurons. The efficacy of inter-neuronal transmission is modulated by the presence of inhibitory synapses, their number and their recruitment.

The consequence of this organization is that the concept of simultaneity, which enables us to freeze within a single slice the multiple elements composing a single event or object in the physical world, does not apply to the functioning of the brain [10]. Once we consider the set of spikes emitted by central cells belonging to the chain of neuronal activity that participates in changing the orientation of the eyes in response to a visual event, they fit within a slice that is much thicker than the slice of spikes emitted by the motor neurons and driving the rotation of the eyes. When it comes to estimating the number of items present in a visual scene as the number of clubs on a playing card, the concept of simultaneity is also implicitly invoked. Some authors have argued that the firing rate of some neurons encodes the number of items on a visual display. However, they have documented neither the duration of the time interval during which the items must aggregate to engender the Gestalt corresponding to the cardinal of the sample, nor the sensitivity of the firing rate to the location of items within the visual field.

Let's consider the case of an orienting response toward the location of a single spot of light briefly flashed in the visual field. Transmitted by each optic nerve, the dual retinal excitation elicits a flow of activity that manifests in the brain as a kind of tree-like explosion. The activity consists of trains of action potentials propagating from the retinal ganglion cells to multiple neurons located in subcortical territories in the diencephalon and brainstem, and from there to various areas of the cerebral cortex [22-23]. The spatiotemporally delimited loci of activity elicited by the transient stimulus on each retina is distributed to multiple groups of neurons whose responses vary in their persistence. As the activity spreads and recruits the groups of cells at the origin of the motor nerves, the diffluence yields to a confluence of activity whose duration will determine the movement amplitude. As the number of spikes emitted by the premotor and motor neurons increases, the contraction of muscle fibers persists and makes the movement longer. The recruitment of additional motor cells and muscle groups enhances the amplitude of the orienting response inasmuch as combining the contraction of extraocular muscles with the contraction of neck muscles yields larger combined eye-head gaze shifts.

However, a visual event should not be reduced to the transmission of signals from the retina to the motor neurons through a multitude of relays. A postural context precedes the occurrence of the visual event. The visual experience depends upon the initial orientation of the eyes and of the head, and thus the sustained contraction of extraocular and neck muscles. When looking at an object located in front of us, its image is projected onto the fovea, from which the central part of the visual field originates. The foveal projection of the object is allowed by appropriate tonic contractions of extra-ocular muscles and neck muscles. A slight deviation of the head toward the right relative to the trunk is compensated by an equivalent orbital deviation of the eyes toward the left. The retinal image remains the same, but the set of contractions and relaxations is different in the extraocular and neck muscles. In other words, in addition to the spontaneous activity of neurons located centrally in the brain, the visual experience combines sensory signals from the retina, proprioceptive signals from the neck muscles and the sustained firing rate that maintains the tonic contraction of muscles and that holds stable the orientation of both eyes and the head. This tonic motor context is far from negligible. Indeed, psychophysical studies have shown that the direction toward which a subject points (with a laser pointer or with one finger) in response to a visual target is altered by unilateral vibration of neck muscles, which provides a false proprioceptive signal [25], during the passive deviation of the eye [26] or more simply while wearing prismatic spectacles. In the latter case and prior to the subsequent visuo-manual adaptation, the target image is projected onto the fovea and the hand aims at a location that is offset with respect to the physical target location.

## 4 Does the firing rate of central neurons encode movement kinematics?

Before explaining why it is unlikely that the firing rate of central neurons encodes a kinematic parameter, we shall first expound the theoretical framework that was used during the last decades to explain the generation of an orienting response to the sudden appearance of a target in the peripheral visual field. This response is characterized by three intervals: a preparatory interval, an interval during which the eyes change their orientation and post-movement interval during which the eyes maintain their orientation toward the target location or nearby.

During the preparatory phase, the eyes do not move while the target-related signals travel across the brain's neuronal networks toward the motor neurons. The stationarity of the eyeballs is assured by the sustained firing rate of motor neurons, which is maintained by a set of tonic premotor activities forming an equilibrium of commands that counterbalance each other. The notion of poly-equilibrium would be more appropriate to name this state of balanced activity because it is not unitary. It involves the channels leading to the generation of saccades, those leading to the generation of slow eye movements, those yielding the near response (accommodation and vergence) and those holding the head [16, 21, 27]. Any suppression of activities within the field of commands participating to this poly-equilibrium alters the direction of gaze while fixating a target. The animal looks toward a location that is offset relative to the target, be it static [28-31] or moving [29-30]. In the cat, the direction of the head can also be offset relative to the location of a food target [28, 34].

The movement phase consists of a phasic contraction of agonist muscles combined with a relaxation of antagonist muscles. The phasic contraction results from a burst of action potentials that motor neurons emit on top of their sustained firing rate, whereas the relaxation is enabled by a pause in their firing rate [21]. By recording the activity of motor neurons and the changes in both muscle tension and eye orientation, Davis-Lopez de Carrizosa and colleagues [35] found that the time course of motoneuronal firing rate better correlates with the time course of muscle tension and eye position than with the time course of eye kinematics (position, velocity and acceleration). Upstream from the motor neurons, the premotor neurons fire with a rate that is quasi-constant, unlike the rise in muscle tension or in eye speed [18]. Intriguingly, some authors claim that the firing rate of neurons located more centrally, in the deep layers of the superior colliculus, correlates with the instantaneous speed of the change in the direction of the line of sight [5, 36]. They support a functional scheme according to which gaze direction is guided toward the target by a negative feedback loop [37-38]. The eye and head premotor neurons would be driven by a motor error command corresponding to the distance between the gaze and target directions. This motor error would result from the continuous comparison between intrinsic signals supposed to encode on the one hand, the gaze-target angle before the movement onset, and on the other hand, the angle of the line of sight as it travels through space. Thus, while gaze moves, the remaining angular distance (motor error) is updated by internal feedback signals encoding the brain's estimate of gaze direction. The saccade would stop once the motor error is zeroed.

The post-movement phase is characterized by a return to sustained activity for maintaining the new posture, i.e., maintaining the eye deviated in the orbits if the head is prevented from moving, and if the head is free to move, maintaining a new gaze direction. Modelling studies have proposed that the sustained activity responsible for the deviation of the eyes in the orbit results from the temporal integration (in the mathematical sense) of the burst of spikes emitted by the premotor neurons. Until Sparks and colleagues showed its periodic character [18], this burst was imagined to convey a signal related to instantaneous eye velocity. Yet, it is worth remembering, as David A. Robinson noted, that "*block diagrams of oculomotor organization serve as a compact description of system behavior but seldom have much bearing on the way in which the real system, composed of nerve and muscle, actually operates. The models thus do not contribute much to the neurophysiology (or neurology) of eye movements and incur the danger of suggesting that there actually are segregated portions of the nervous system which perform the differentiation, integration and other operations indicated in the boxes of the diagrams*" [39].

These models proposed a conceptual framework that was fruitful insofar as they provoked debates. Their predictions stimulated the undertaking of numerous experiments, which brought new observations. While many results were compatible with the models, others were difficult to explain and barely discussed. More fundamentally, if the signals transmitted between the modules have no neurophysiological substrate, then the model becomes a scientifically embarrassing fiction insofar as it is irrefutable; for it is impossible to demonstrate that a fiction has no real counterpart.

We shall now examine the evidence supporting the claim that eye velocity is encoded in the firing rate of neurons or in the activity of a population of neurons. Based on empirical data collected during difficult experiments, the evidence remains as weak as that put forward by other authors to defend the idea of a coding of space, of time and quantity in the brain activity.

### 4.1 The case of Purkinje cells in the oculomotor vermis

In 2015, Herzfeld and colleagues [40] reported a relation between the saccade speed and the population activity of one category of cells in the cerebellar vermis, the Purkinje cells. However, the close examination of their figures leads one to the conclusion that this relation is an artefact resulting from non-homogeneous measurements. The comparison between their figure 3b (which plots the speed of all recorded saccades and the corresponding neural activity) and figure 3f (which plots the speed of a subset of saccades, those with an amplitude of 10 deg) teaches us that the 10 deg saccades with high-velocity were extremely rare. Figure 3b (bottom) shows that the average change in the population response during 10 deg saccades (red trace) is approximately 950 spikes/s whereas figure 3f (bottom) shows that the approximate changes are 2450 spikes/s for saccades with 650°/s peak speed (blue trace), 1500 spikes/s for saccades with 525°/s peak speed (green trace), and 750 spikes/s for saccades with 400°/s peak speed (red traces). A grand average value of 950 spikes/s (Fig. 3b) can be obtained if the saccades with 650, 525 and 400°/s peak speed (Fig. 3f) correspond to 5%, 15% and 80% of 10 deg saccades, respectively. These percentage values indicate that the distribution of peak speed values was strongly asymmetric, which is confirmed when figure 3f is compared to figure 1A of their supplementary material. Therein, the average peak speed of 10 deg saccades is approximately 440 deg/s, a value at the bottom of peak values reported in figure 3f. Consequently the reported correlation between the population activity and saccade speed results from an asymmetric distribution of values, which can be explained by the fact that the fastest saccades were recorded in one single monkey (possibly in the monkey W; see Fig. 3 in [39]) and not in the other monkeys. In other words, the relation between saccade speed and population activity could merely result from comparing different subjects. For example, when we examine the data of Kojima and colleagues [41], no saccade of 10 deg amplitude has a peak speed higher than 500 deg/s in monkey B (their Figure 3A) in contrast to the findings for monkey W (their Fig. 3D). That different monkeys produce the same saccade amplitude with a different peak velocity is well-known (see also Fig. 1 in [42]).

Another concern in the study of Herzfeld and colleagues is that the correlation between saccade speed and population activity assumed an equal number of "bursting" and "pausing" cells. Indeed, the majority of studies report a numerical difference. McElligott and Keller [43] reported 71% "bursting" and 29% "pausing" cells, Ohtsuka & Noda [42] 71% "bursting" and 18% "pausing" cells and Helmchen & Büttner [45] 67% "bursting" and 10% "pausing" cells. Herzfeld and colleagues [40] acknowledged that the population response could not predict the real-time saccade velocity if each cluster was not composed of roughly equal numbers of pausing and bursting cells. They also indicated that when 70% of the population were bursting neurons, the peak response still scaled linearly with saccade peak velocity. However, considering our primary concern (combining data collected in different monkeys), the correlation might disappear if the data collected in the non-ordinary monkey (W) were removed from the population. In summary, the suggestion that the activity of the population of Purkinje cells predicts "eye velocity in real-time" is weakly founded.

A relation between the population response of Purkinje cells and the saccade speed is also not compatible with the electrophysiological recordings in the caudal fastigial nucleus (cFN), the nucleus that houses the post-synaptic neurons targeted by Purkinje cells in the vermis. Indeed, the majority of studies report no modulation of saccade-related bursts with respect to saccade speed in cFN (see [46] for a detailed review of the literature). Only one study has suggested a relation between cFN activity and saccade velocity [47]. However, even this relationship may be due to different levels of alertness between the early and late parts of the recording sessions, which were long and boring. Fuchs and colleagues indeed reported an influence of alertness on the firing rate of cFN neurons [48].

In conclusion, the fact that unilateral and brief intrasaccadic changes of activity in the vermis or in the cFN impair the trajectory of saccades to a visual target [49-51] does not necessarily imply that the instantaneous firing rate of neurons encodes the instantaneous saccade velocity. It merely tells us that, through their bilateral projections to the premotor burst neurons, the spikes emitted by neurons in the left and right cFN contribute to the phasic contraction of agonist muscle fibers and to the phasic relaxation of antagonist muscle fibers [52-53].

**4.2 The case of saccade-related burst neurons in the deep superior colliculus**

Following a study showing in the monkey that muscimol injection in the deep layers of the superior colliculus (SC) altered the velocity of saccades toward a peripheral target [54], the suggestion was made by Berthoz and colleagues [55] that the firing rate of neurons in the deep superior colliculus encode the velocity of a saccade. A resemblance was reported between the instantaneous firing rate of seven neurons and the instantaneous speed of saccades produced by one cat. To support their argument, they showed three recording samples illustrating this resemblance. Unfortunately, they reported neither the number of times this resemblance was tested nor the proportion of neurons exhibiting a relation between the firing rate and saccade velocity. Reporting that a neuron emits a burst of action potentials with a time course that roughly resembles the saccade velocity one time out of ten recordings does not have the same meaning as reporting that this resemblance appears for each saccade. Likewise, reporting that seven out of ten neurons emit a burst of spikes with a time course resembling the saccade velocity profile does not have the same meaning as reporting that these neurons represent one percent of SC neurons. In other words, the alleged properties are not supported by a sufficient sampling of neurons.

The suggestion for a coding of instantaneous velocity by the instantaneous firing rate of collicular neurons was taken up by electrophysiological studies in macaque monkeys who were tested with the head restrained [36, 56] or unrestrained [5]. According to van Opstal, the "*SC population activity encodes the instantaneous kinematics of the desired gaze shift through its firing rates, whereas the gaze-shift amplitude is encoded by the number of spikes in the saccade-related burst*" [5]. Several figures reported in his study lead to the impression that all saccade-related neurons in the SC cease emitting spikes shortly before the gaze shift ends. However, this inference may result from the author's choice of example neurons because all saccade-related neurons in the SC do not exhibit such an abrupt cessation of activity shortly before the gaze saccade ends [57]. In the head-restrained monkey, numerous studies have documented neural discharges that persist beyond saccade end [58-64]. Moreover, if the time course of the instantaneous spike density and the time course of "gaze-track" velocity look remarkably similar for the neuron illustrated in Fig. 9A-B, it is because the spikes that preceded the saccade-related burst were removed. The graph at the bottom of Fig. 9A shows that twenty milliseconds before gaze change onset, the firing rate rises from 0 to 200 spikes per second. Examination of Fig. 4A reveals that such an enhancement (from zero to 200 spikes/s) is rare for the same neuron. Presumably, the spikes emitted before the onset of an analysis interval (from 20 ms before gaze onset to 20 ms before gaze end) were removed. However, most studies have reported collicular saccade-related neurons with a long prelude of activity before saccade onset. By removing the spikes emitted before saccade onset and after saccade end, the spike density function exhibits a rising phase followed by a decline as the accelerating and decelerating parts of the velocity. Thus, the correlation between the instantaneous firing rate and gaze velocity is the artificial consequence of selecting a portion of the neuron's activity. It does not support a causal relation either.

Presumably, van Opstal selected the spikes emitted from 20 ms before gaze onset to 20 ms before gaze shift end because they happened in the time interval during which the so-called omnipause neurons cease firing. Located in the nucleus raphe interpositus, these neurons are known to inhibit the premotor burst neurons. However, if the gaze-shift amplitude is encoded by the number of spikes in the saccade-related burst, then experimentally increasing the pause duration should increase the number of action potentials and lead to hypermetric saccades. Empirical data do not support this prediction insofar as a lesion of nucleus raphe interpositus does not lead to dysmetric saccades [65-66]. Furthermore, by cooling the frontal eye fields and recording the firing rate of saccade-related burst neurons in the SC, Peel and colleagues found that the total number of spikes was reduced throughout the entire response field whereas the saccade amplitude remained unchanged [67]. Their results refute the conjecture of an encoding of gaze-shift amplitude by the number of spikes in the saccade-related burst, even during visually-guided trials (their Fig. 1).

Finally, the evidence brought by Smalianchuk et al. [36] and by van Opstal [5] rests upon the replacement of each spike by a Gaussian curve. By transforming the sequence of action potentials into a continuous curve, the correlation can be tested. However, we shall explain that establishing a continuity in the firing rate of neurons is potentially misleading if the parameter critical in neural transmission is not the sequence of spikes emitted by the recorded neuron, but is instead the membrane potentials of post-synaptic neurons and the synchrony of action potentials bombarding them.

Before this explanation, we shall stress the fact that the premotor burst neurons targeted by the SC neurons exhibit firing rates whose time course is not correlated with the instantaneous saccade velocity. Their instantaneous firing rate is almost constant; their intersaccadic interval is quasi-constant, like a clock [18]. While the firing rate of motor neurons can be correlated with the instantaneous saccade velocity because of the causal role in the contraction of extraocular muscles, the correlation disappears when we study the firing rate of their presynaptic input from the premotor burst neurons. The discovery that a model factoring muscle tension and its first derivative fits the firing rate of motor neurons better than a model factoring eye position and its first and second derivatives (velocity and acceleration) [35] brings more difficulty. Indeed, if the activity of SC neurons encoded the instantaneous velocity of the saccade, then, downstream, this kinematic command must be transformed into a dynamic (force-related) command. Any attempt at an explanation might be vain because the time course of the change in eye orientation does not match with the time course of the change in muscle tension. The decrease in muscle force starts before saccade end with a lead time (between the onset of the decay and the end of the saccade) which is longer with larger amplitudes [68]. Thus, the same physical phenomena, saccades, exhibit different time courses depending upon the measurement method, i.e., depending upon whether one records the change in the orientation of the eye or the change in muscle tension.

**4.3 On the possibility of encoding kinematics or dynamics by central neurons**

We shall now explain why it is dubious that the firing rate of central neurons encodes the time course of a kinematic (or kinetic) parameter. Our explanation rests upon the anatomo-physiological fact that any neuron receives action potentials emitted by presynaptic neurons that are distributed in several brain regions, more or less distant from the recorded neuron. It also rests upon the fact that the probability of emitting an action potential depends upon the relative synchrony of pre-synaptic action potentials [69].

If we consider the firing rate of motor neurons that innervate the extraocular muscles, a correlation between the saccade velocity and the sequence of action potentials is very likely because the latter cause the contraction of extraocular muscle fibers, which in turn exerts the torque responsible for the rotation of the eyeball [70]. However, if we now turn to the premotor neurons innervating the motor neurons involved, for example, in horizontal saccades, interpreting the correlation is complicated by the fact that several inputs converge onto the motor neurons. They receive inputs from excitatory burst neurons located in the ipsilateral paramedian reticular formation [71], inhibitory burst neurons in the contralateral medullary reticular formation [72] and from burst-tonic neurons located bilaterally in the left and right nuclei prepositus hypoglossi (NPH) and medial vestibular nuclei (MVN) [11-13, 16, 73]. The discharge patterns of these different groups of neurons do not exhibit identical time courses. Consequently, since the pre-synaptic inputs to the motor neurons originate from neurons distributed in different groups, the correlation between the firing rate and the saccade kinematics becomes weaker. Inevitably, the correlation becomes even weaker when we study the firing rate of neurons that project to those premotor neurons, directly like neurons in the caudal fastigial nuclei [47-48] or superior colliculi [74].

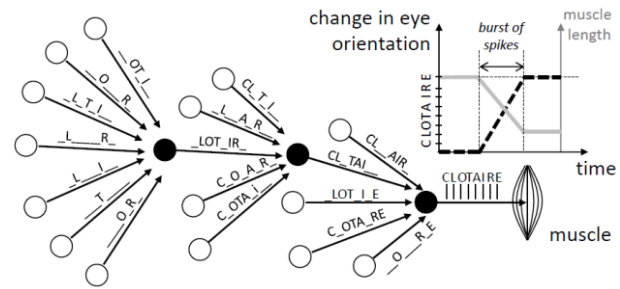

**Fig. 1:** Black-colored cells emit a "letter-spike" (C, L, O, T, A, I, R and E) in response to three simultaneous presynaptic "letter-spikes".

Let's consider the successive values of eye orientation that make a saccade of 8 degrees amplitude starting from straight ahead. The values are "1-2-3-4-5-6-7-8"; they are ordinal. As explained at the beginning of our article, each value is chained to its preceding and following values by an infinite number of intermediate values. The sequence "**4**-**3**-**7**-**2**-**1**-**8**-**7**-**5**" is not a physically plausible movement. To be a movement, it would have to be "1-2-3-**4**-**3**-4-5-6-**7**-6-5-4-3-**2**-**1**-2-3-4-5-6-7-**8**-**7**-6-**5**. Let's now associate the sequence of numerical values "1-2-3-4-5-6-7-8" with the sequence of letters in the given name "C-L-O-T-A-I-R-E" and imagine that each spike emitted by a motor neuron contracts the muscle fibers and increases the eye orientation. The first spike ("C") is responsible for increasing the eye orientation from "0" to "1", the second ("L") for increasing it from "1" to "2", the third from "2" to "3" etc. As stated previously, the probability of emitting one spike depends upon the synchrony of pre-synaptic action potentials. Then let's call "C" all the pre-synaptic simultaneous spikes that are responsible for the post-synaptic spike "C", "L" all the pre-synaptic simultaneous spikes responsible for the post-synaptic spike "L", etc. As we move our investigation from the motor neurons to more central neurons, Fig. 1 shows that the sequence of letters "C-L-O-T-A-I-R-E" is absent from the set of presynaptic sequences. Since each sequence of letters corresponds to a sequence of ordered numerical values, none of them corresponds to any physical measurement. Finally, because of the distributed character of the brain connectivity, different sequences of letters (different sequences of spikes) can lead to kinematically identical saccades. Thus, identical saccades can be associated with variable firing rates.

## Conclusion

A few years after the publication of *The Origin of Species* by Darwin, Spencer defended the thesis according to which the *a priori* forms of intelligence such as space and time are a heritage of animal evolution that has been registered in the physiological organization of the nervous system. "*Spatial relationships have been the same, not only for all men, all primates and all orders of mammals from which we descend, but also for all orders of less elevated beings*". They are "*expressed in defined nervous structures, congenitally constituted to act in a certain way, and incapable of acting in a different way*" [75]. One hundred and forty years later, Dehaene and Brannon put forth the same thesis: "*in the course of their evolution, humans and many other animal species have internalized basic codes and operations isomorphic to the physical and arithmetic laws that govern the interaction of objects in the external world. Indeed, there is now considerable evidence that space, time and number are part of the essential toolkit that adult humans share with infants and with many nonhuman animals*"[1]. Likewise, Berthoz contended that the laws of Newtonian physics are internalized in the brain functioning [76] and that mathematics are "*the expression of laws and mechanisms which are in operation in the brain ... a language of the brain on the brain itself, that is to say, since the brain is part of the physical world,* [a language] *about the Universe and its laws*" [77]. An isomorphic relation between the patterns of nature and the patterns of cognition is a central thesis in the domain of evolutionary epistemology [78]

Against this internalization of the laws of nature in the structure of organisms, we defend a more active and contrarian notion of intelligence and living forms. Cognition, physiology and even morphology are not merely representations, embodied copies of the external world. They are counter-reactions, explorations of possibilities for expansion, attempts to trace sustainable paths through an environment characterized by constraints specific to each level of complexity [79]. One may be tempted to claim that "*the philosophy of mathematics has not paid much attention to the biological context of mathematics, although it goes without saying that mathematicians are biological organisms and that they produce mathematical theories and practices with their brains*" [4]. However, by reducing the mind to the brain and intellectual development to brain maturation, we confound the brain functioning of an individual with the historical and collective evolution of scientific discourse, which is stored in books, libraries and technical artefacts, and obeys Boolean logic. "*Mathematics is a collective work of art that derives its objectivity through social interactions*" [7] and "*the locus of scientific rationality is not the individual scientist but the scientific community*" [80]. As early as in 1909, Emile Durkheim explained that the categories of human thought are not *"a priori, universal ..., locked in the physiology or biology of Homo sapiens, but socio-historical…"* [81], that they are *"like learned instruments of thought, which human groups have laboriously forged over the centuries and where they have accumulated the best of their intellectual capital*" [82]. By placing *"in man (and more precisely in his brain) what is the mere product of social relationships maintained with other men and objects (products of human activity)"* [80], some cognitive neuroscientists have diverted the search for core networks that make the complexity of the brain [79], overlooking also the human collective and socio-historical dimension of mathematics.

## Acknowledgments

The author thanks the reviewers for detailed comments that significantly improved the text. This article is dedicated to the memory of André Goffart.